\documentclass[journal]{IEEEtran}
\usepackage{float}
\usepackage{cite}										
\usepackage{amsmath}									
\usepackage{mathtools}
\usepackage{graphicx}

\usepackage{epstopdf}
\usepackage{float}
\usepackage{bm,upgreek}
\usepackage{amsfonts}
\usepackage{enumitem}
\usepackage{pgfplots}
\pgfplotsset{compat=1.5}
\emergencystretch=\maxdimen
\allowdisplaybreaks
\ifCLASSINFOpdf
\else
\fi

\ifCLASSOPTIONcompsoc
  \usepackage[caption=false,font=normalsize,labelfont=sf,textfont=sf]{subfig}
\else
 \usepackage[caption=false,font=footnotesize]{subfig}
\fi

\begin{document}
\title{Fast Real-Time DC State Estimation in Electric Power Systems Using Belief Propagation}

\author{Mirsad~Cosovic,~\IEEEmembership{Student Member,~IEEE,}
        Dejan Vukobratovic,~\IEEEmembership{Member,~IEEE}

\thanks{M. Cosovic is with Schneider Electric DMS NS, Novi Sad, Serbia (e-mail: mirsad.cosovic@schneider-electric-dms.com). D. Vukobratovic is with Department of Power, Electronic and Communications Engineering, University of Novi Sad, Novi Sad, Serbia (e-mail: dejanv@uns.ac.rs). Demo source code available online at https://github.com/mcosovic.}

\thanks{This paper has received funding from the EU 7th Framework Programme for research, technological development and demonstration under grant agreement no. 607774.}}

\maketitle

\begin{abstract}
We propose a fast real-time state estimator based on the belief propagation algorithm for the power system state estimation. The proposed estimator is easy to distribute and parallelize, thus alleviating computational limitations and allowing for processing measurements in real time. The presented algorithm may run as a continuous process, with each new measurement being seamlessly processed by the distributed state estimator. In contrast to the matrix-based state estimation methods, the belief propagation approach is robust to ill-conditioned scenarios caused by significant differences between measurement variances, thus resulting in a solution that eliminates observability analysis. Using the DC model, we numerically demonstrate the performance of the state estimator in a realistic real-time system model with asynchronous measurements. We note that the extension to the AC state estimation is possible within the same framework.
\end{abstract}


\IEEEpeerreviewmaketitle

\section{Introduction}
The state estimation (SE) function is a part of the energy management system that allows for monitoring of electric power systems. Input data for the SE arrive from supervisory control and data acquisition (SCADA) technology. SCADA provides communication infrastructure to collect legacy measurements (voltage and line current magnitude, power flow and injection measurements) from measurement devices and transfer them to a central computational unit for processing and storage. In the last decades, phasor measurement units (PMUs) were developed that measure voltage and line current phasors and provide highly accurate measurements with high sampling rates. PMUs were instrumental to the development of the wide area measurement systems (WAMSs) that should provide real-time monitoring and control of electric power systems \cite{zhu, scaglione}. The WAMS requires significant investments in deployment of a large number of PMUs across the system, which is why SCADA systems will remain important technology, particularly at medium and low voltage levels. 
 
Monitoring and control capability of the system strongly depends on the SE accuracy as well as the periodicity of evaluation of state estimates. Ideally, in the presence of both legacy and PMU measurements, SE should run at the scanning rate (seconds), but due to the computational limitations, practical SE algorithms run every few minutes or when a significant change occurs \cite{monticelli}. In this work, we propose a fast real-time state estimator based on the belief propagation (BP) algorithm. Using the BP, it is possible to estimate state variables in a distributed fashion. In other words, unlike the usual scenario where measurements are transmitted directly to the control center, in the BP framework, measurements are locally collected and processed by local modules (at substations, generators or load units) that exchange BP messages with neighboring local modules. Furthermore, even in the scenario where measurements are transmitted to the centralized control entity, the BP solution is advantageous over the classical centralized solutions in that it can be easily distributed and parallelized for high performance. 

Compared to our recent work on BP-based SE \cite{tsp_cosovic, smartgrid} that addresses classical (static) SE problem, this paper is an extension to the real-time model that operates continuously and accepts asynchronous measurements from different measurement subsystems. More precisely, we assume presence of both SCADA and WAMS infrastructure, and without loss of generality, we observe active power flow and injection measurements (from SCADA), and voltage phase angle measurements (from WAMS). We present appropriate models for measurement arrival processes and for the process of measurement deterioration (or ``aging'') over time. Such measurements are continuously integrated into the running instances of distributed BP-based modules. For simplicity, we present the real-time BP-based SE applied on the DC SE model, while extension to the AC SE model follows similar lines as in the static SE scenario \cite{tsp_cosovic}. Our extensive numerical experiments on the example IEEE 14 system show that the BP algorithm is able to provide real-time SE performance. Furthermore, the BP-based SE is robust to ill-conditioned systems in which significant difference arise between measurement variances, thus allowing state estimator that runs without observability analysis. Note that in this paper, we do not address the convergence guarantees for the BP-based solution \cite{fu}, and we leave the detailed treatment of convergence for our future work.

The structure of this paper is as follows: In Section II, we provide background on conventional and BP-based SE. Section III described the proposed fast real-time BP-based SE, while Section IV considers the performance and numerical results for the IEEE 14 bus test case. Concluding remarks are provided in Section V. 

\section{Background}
\subsection{State Estimation in Electric Power Systems}
The main SE routines comprise the SE algorithm, network topology processor, observability analysis and bad data analysis. The core of the SE is \emph{the SE algorithm} which provides a state estimate of the system, i.e., the set of all complex bus voltages, based on the network topology and set of measurements $\mathcal{M}$. Using information about switch and circuit breaker positions \emph{the network topology processor} generates a bus/branch model of the power network and assigns real-time measurement devices (legacy and/or PMU devices) across the bus/branch model \cite[Sec.~1.3]{abur}. As a result, the graph $\mathcal{G} =$ $(\mathcal{V},\mathcal{E})$ representing the power network is defined, where the set of nodes $\mathcal{V} =$ $\{1,\dots,n \}$ represents the set of buses, while the set of edges $\mathcal{E} \subseteq \mathcal{V} \times \mathcal{V}$ represents the set of branches. In addition, the set of real-time measurements $\mathcal{M}_{\mathrm{rt}} \subseteq \mathcal{M}$ is connected to the graph $\mathcal{G}$.

According to the location and the type of real-time measurements \emph{the observability analysis} determines observable and unobservable islands. Within the observable islands, it is possible to obtain unique state estimates from the available set of real-time measurements $\mathcal{M}_{\mathrm{rt}}$, which is not the case within unobservable parts of the system. Once observability analysis is done, pseudo-measurements can be added, in order for the entire system to be observable \cite[Sec.~4.6]{abur}, \cite{monticelli}. The set of pseudo-measurements $\mathcal{M}_{\mathrm{ps}} \subset \mathcal{M}$ represents certain prior knowledge (e.g., historical data) of different electrical quantities and they are  usually assigned high values of variances \cite[Sec.~1.3]{abur}. As detailed later, we assume that, at a given time, the system measurements are either real-time or pseudo-measurements, i.e., the sets $\mathcal{M}_{\mathrm{rt}}$ and $\mathcal{M}_{\mathrm{ps}}$ are disjoint $\mathcal{M}_{\mathrm{rt}} \cap \mathcal{M}_{\mathrm{ps}} = \emptyset$ and their union is the set $\mathcal{M} = \mathcal{M}_{\mathrm{rt}} \cup \mathcal{M}_{\mathrm{ps}}$.   

The observability analysis provides the measurement model which can be described as the system of equations:
		\begin{equation}
        \begin{aligned}
        \mathbf{z}=\mathbf{h}(\mathbf{x})+\mathbf{u},
        \end{aligned}
		\label{SE_model}
		\end{equation}
where $\mathbf {x}=(x_1,\dots,x_{n})$ is the vector of state variables, $\mathbf{h}(\mathbf{x})=$ $(h_1(\mathbf{x})$, $\dots$, $h_k(\mathbf{x}))$ is the vector of measurement functions,  $\mathbf{z} = (z_1,\dots,z_k)$ is the vector of independent measurement values, and $\mathbf{u} = (u_1,\dots,u_k)$ is the vector of uncorrelated measurement errors. The SE problem is commonly an overdetermined system of equations $(k>n)$ usually defined by both real-time measurements and pseudo-measurements \cite[Sec.~2.1]{monticelliBook}.

Each measurement $M_i \in \mathcal{M}$ is associated with measured value $z_i$, measurement error  $u_i$ and measurement function $h_i(\mathbf{x})$. Under the assumption that measurement errors $u_i$ follow a zero-mean Gaussian distribution, the probability density function associated with the measurement $M_i$ equals:
		\begin{equation}
        \begin{gathered}
        \mathcal{N}(z_i|\mathbf{x},\sigma_i^2) = \cfrac{1}{\sqrt{2\pi \sigma_i^2}} 
        \exp\Bigg\{\cfrac{[z_i-h_i(\mathbf{x})]^2}{2\sigma_i^2}\Bigg\},
        \end{gathered}
		\label{SE_Gauss_mth}
		\end{equation}
where $\sigma_i^2$ is the variance of the measurement error  $u_i$, and the measurement function $h_i(\mathbf{x})$ connects the vector of state variables $\mathbf{x}$ to the value of the measurement $M_i$.

One can find the state estimate $\hat{\mathbf x}$ via maximization of the likelihood function $\mathcal{L}(\mathbf{z}|\mathbf{x})$, which is defined via likelihoods of $k$ independent measurements:  
		\begin{equation}
        \begin{gathered}
		\hat{\mathbf x}=
		\mathrm{arg} \max_{\mathbf{x}}\mathcal{L}(\mathbf{z}|\mathbf{x})=
		\mathrm{arg} \max_{\mathbf{x}}  
		\prod_{i=1}^k \mathcal{N}(z_i|\mathbf{x},\sigma_i^2).
        \end{gathered}
		\label{SE_likelihood}
		\end{equation}

It can be shown that the solution of \eqref{SE_likelihood} can be obtained by solving the following optimization problem, known as the weighted least squares (WLS) problem \cite[Sec.~9.3]{wood}:
		\begin{equation}
        \begin{gathered}
		\hat{\mathbf x} =
		\mathrm{arg}\min_{\mathbf{x}} \sum_{i=1}^k 
		\cfrac{[z_i-h_i(\mathbf x)]^2}{\sigma_i^2}.
        \end{gathered}
		\label{SE_WLS_problem}
		\end{equation}
The state estimate $\hat{\mathbf x}$ that represents the solution of the optimization problem \eqref{SE_WLS_problem} is known as the WLS estimator and it is identical to the maximum likelihood solution.  

\subsection{DC State Estimation}
The DC model is dealing with linear measurement functions $\mathbf{h}(\mathbf{x})$ and it is obtained by linearisation of the AC model \cite{tsp_cosovic}. Therefore, the DC SE takes only bus voltage angles $\mathbf x  \equiv \bm \uptheta$ as state variables and the set of measurements $\mathcal{M}$ involves the active power flow at the branch  $(i,j) \in \mathcal{E}$, the active power injection into the bus $i \in \mathcal{V}$ and the bus voltage angle at the bus $i \in \mathcal{V}$, with measurement functions defined as follows:  
		\begin{subequations}
        \begin{align}
         h_{P_{ij}}(\cdot) &= 
        -b_{ij}(\theta_{i}-\theta_{j})
		\label{DC_active_flow}\\
        h_{P_{i}}(\cdot) &=
        \sum_{j \in \mathcal{H}_i \setminus i} h_{P_{ij}}(\cdot)
        \label{DC_active_injection}\\
         h_{\theta_{i}}(\cdot) &=
        \theta_i,
        \label{DC_angle}
        \end{align}
		\label{DC_meas_functions}%
		\end{subequations}
where $\theta_i$ and $\theta_j$ are bus voltage angles at buses $i$ and $j$, $b_{ij}$ is susceptance of the branch and $\mathcal{H}_i\setminus i$ is the set of buses incident to the bus $i$.

The DC state estimate $\hat{\mathbf x} \equiv \hat{\bm \uptheta}$, which is a solution to the WLS problem \eqref{SE_WLS_problem}, is obtained through non-iterative procedure by solving the system of linear equations:  
		\begin{equation}
        \begin{aligned}  
		\big(\mathbf H^\mathrm{T} \mathbf W \mathbf H \big) \hat{\mathbf x} =		
		\mathbf H^\mathrm{T} \mathbf W\mathbf z, \label{DC_WLS}\\
		\end{aligned}
        \end{equation}
where $\mathbf{H}\in \mathbb {R}^{k \times n}$ is the Jacobian matrix of measurement functions \eqref{DC_meas_functions}, and $\mathbf{W}\in \mathbb {R}^{k \times k}$ is a diagonal matrix containing inverses of measurement variances.

\subsection{Factor Graphs and Belief Propagation Algorithm}
\textbf{Factor graph construction:} The factor graph describes a factorization of the likelihood function $\mathcal{L}(\mathbf{z}|\mathbf{x})$. It comprises the set of factor nodes $\mathcal{F}$ and the set of variable nodes $\mathcal{X}$. In the DC scenario, the vector of state variables $\bm \uptheta$ determines the set of variable nodes $\mathcal{X} = \{\theta_1,\dots,\theta_n\}$, while the set of measurements $\mathcal{M}$ defines the set of factor nodes $\mathcal{F} =$ $\{f_1,\dots,f_k\}$. Each measurements defines a corresponding factor $\mathcal{N}(z_i|\mathbf{x},\sigma_i^2)$ of the likelihood function which is represented by a factor node. A factor node $f_i$ connects to a variable node $x_s \in \mathcal{X}$ if and only if the state variable $x_s$ is an argument of the corresponding measurement function $h_i(\mathbf x)$ \cite{tsp_cosovic}.

\textbf{Belief propagation algorithm:} The BP algorithm efficiently calculates marginal distributions of state variables by passing two types of messages along the edges of the factor graph: i) a variable node to a factor node, and ii) a factor node to a variable node messages. Both variable and factor nodes in a factor graph process the incoming messages and calculate outgoing messages. The marginal inference provides marginal probability distributions that is used to estimate values $\hat{\mathbf x}$ of state variables $\mathbf x$. Next, we describe a version of BP algorithm called Gaussian BP, where all the messages represent Gaussian distributions. 

\begin{figure}[ht]
	\centering
	\begin{tabular}{@{}c@{}}
	\subfloat[]{\label{Fig_vf}
	\includegraphics[width=3.7cm]{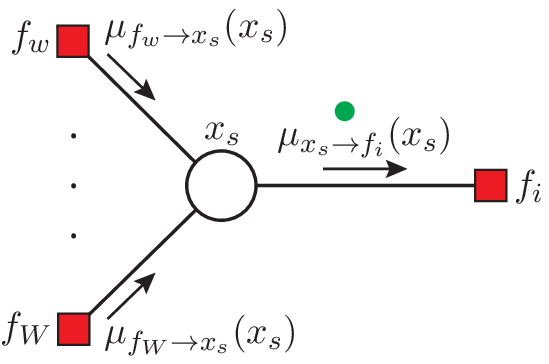}}
	\end{tabular}\quad
	\begin{tabular}{@{}c@{}}
	\subfloat[]{\label{Fig_fv}
	\includegraphics[width=3.7cm]{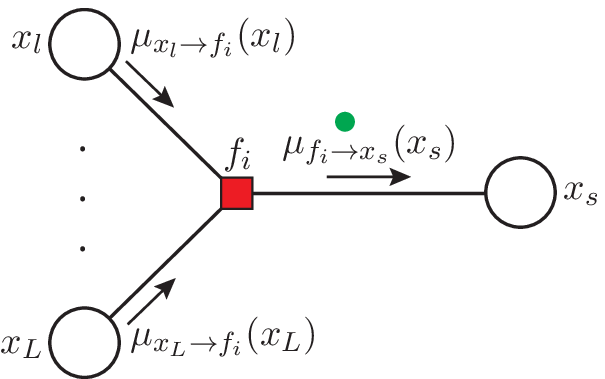}}
	\end{tabular}\\
	\begin{tabular}{@{}c@{}}
	\subfloat[]{\label{Fig_marginal}
	\includegraphics[width=3.7cm]{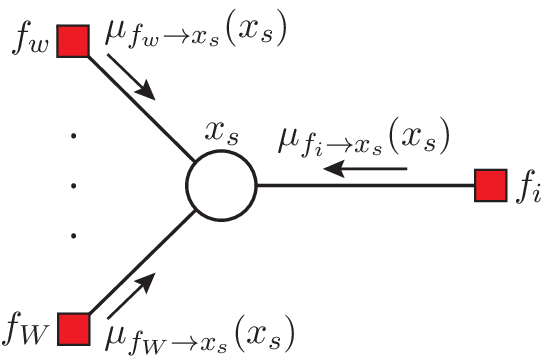}} 
	\end{tabular}
	\caption{Message $\mu_{x_s \to f_i}(x_s)$ from variable node $x_s$ to factor node $f_i$ 
	(subfigure a), message $\mu_{f_i \to x_s}(x_s)$ from factor node $f_i$ to variable node 
	$x_s$ and marginal inference of the variable node $x_s$ (subfigure c).}
	\label{Fig_ex_DC}
	\end{figure}

\emph{Message from a variable node to a factor node:} Consider a part of a factor graph shown in Fig. \ref{Fig_vf} with a group of factor nodes $\mathcal{F}_s=\{f_i,f_w,...,f_W\}$ $\subseteq$ $\mathcal{F}$ that are neighbours of the variable node $x_s$ $\in$ $\mathcal{X}$. Let us assume, for the time being, that the incoming messages $\mu_{f_w \to x_s}(x_s)$, $\dots$, $\mu_{f_W \to x_s}(x_s)$ into the variable node $x_s$ are Gaussian and represented by their mean-variance pairs $(z_{f_w \to x_s},\sigma_{f_w \to x_s}^2)$, $\dots$, $(z_{f_W \to x_s},\sigma_{f_W \to x_s}^2)$.
	 \noindent	

It can be shown that the message $\mu_{x_s \to f_i}(x_s)$ from the variable node $x_s$ to the factor node $f_i$ is proportional (i.e. $\propto$) to the Gaussian function:  
		\begin{equation}
        \begin{aligned}
		\mu_{x_s \to f_i}(x_s) 
		\propto 
		\mathcal{N}(z_{x_s \to f_i}|x_s,
		\sigma_{x_s \to f_i}^2),		
        \end{aligned}
		\label{BP_Gauss_vf} 
        \end{equation}
with mean $z_{x_s \to f_i}$ and variance $\sigma_{x_s \to f_i}^2$ obtained as: 
		\begin{subequations}
        \begin{align}
        z_{x_s \to f_i} &= 
        \Bigg( \sum_{f_a \in \mathcal{F}_s\setminus f_i}
        \cfrac{z_{f_a \to x_s}}{\sigma_{f_a \to x_s}^2}\Bigg)
        \sigma_{x_s \to f_i}^2
        \label{BP_vf_mean}\\
		\cfrac{1}{\sigma_{x_s \to f_i}^2} &= 
		\sum_{f_a \in \mathcal{F}_s\setminus f_i}
		\cfrac{1}{\sigma_{f_a \to x_s}^2}.
		\label{BP_vf_var}
        \end{align}
		\label{BP_vf_mean_var}%
		\end{subequations}   

After the variable node $x_s$ receives the messages from all of the neighbouring factor nodes from the set $\mathcal{F}_s\setminus f_i$, it evaluates the message $\mu_{x_s \to f_i}(x_s)$ according to \eqref{BP_vf_mean_var} and sends it to the factor node $f_i$. 

\emph{Message from a factor node to a variable node:} Consider a part of a factor graph shown in Fig. \ref{Fig_fv} that consists of a group of variable nodes $\mathcal{X}_i = \{x_s, x_l,...,x_L\}$ $\subseteq$ $\mathcal X$ that are neighbours of the factor node $f_i$ $\in$ $\mathcal{F}$. The message $\mu_{f_i \to x_s}(x_s)$ can be computed only when all other incoming messages (variable to factor node messages) are known. Let us assume that the messages into factor nodes are Gaussian, denoted by: 
		\begin{equation}
        \begin{aligned}
		\mu_{x_l \to f_i}(x_l) &\propto
		\mathcal{N}(z_{x_l \to f_i}|x_l,
		\sigma_{x_l \to f_i}^2)\\
		& \vdotswithin{\propto}\\ 
		\mu_{x_L \to f_i}(x_L) &\propto
		\mathcal{N}(z_{x_L \to f_i}|x_L,
		\sigma_{x_L \to f_i}^2).
        \end{aligned}
		\label{BP_incoming_vf}
		\end{equation} 
The Gaussian function associated with the factor node $f_i$ is given by:
		\begin{multline}
		\mathcal{N}(z_i|x_s,x_l,\dots,
		x_L, \sigma_i^2)\\ 
		\propto 
        \exp\Bigg\{\cfrac{[z_i-h_i
        (x_s,x_l,\dots,x_L)]^2}
        {2\sigma_i^2}\Bigg\}.
		\label{BP_Gauss_measurement_fun}
		\end{multline}
The linear function $h_i(x_s,x_l,\dots,x_L)$ can be represented in a general form as:
		\begin{equation}
        \begin{gathered}
		h_i(x_s,x_l,\dots,x_L) =
		C_{x_s} x_s + 
		\sum_{x_b \in \mathcal{X}_i\setminus x_s} 
		C_{x_b} x_b,
		\end{gathered}
		\label{BP_general_measurment_fun}
		\end{equation}
where $\mathcal{X}_i\setminus x_s$ is the set of variable nodes incident to the factor node $f_i$, excluding the variable node $x_s$. 

It can be shown that the message $\mu_{f_i \to x_s}(x_s)$ from the factor node $f_i$ to the variable node $x_s$ is represented by the Gaussian function:
		\begin{equation}
        \begin{aligned}
		\mu_{f_i \to x_s}(x_s) \propto 
		\mathcal{N}(z_{f_i \to x_s}|
		x_s,\sigma_{f_i \to x_s}^2),
        \end{aligned}
		\label{BP_Gauss_fv}
		\end{equation}
with mean $z_{f_i \to x_s}$ and variance $\sigma_{f_i \to x_s}^2$ obtained as:
		\begin{subequations}
        \begin{align}
		z_{f_i \to x_s} &=         
        \cfrac{1}{C_{x_s}} \Bigg(z_i -  
        \sum_{x_b \in \mathcal{X}_i \setminus x_s} 
        C_{x_b} z_{x_b \to f_i}  \Bigg)
        \label{BP_fv_mean}\\
        \sigma_{f_i \to x_s}^2 &=         
        \cfrac{1}{C_{x_s}^2} \Bigg( \sigma_i^2 +  
        \sum_{x_b \in \mathcal{X}_i \setminus x_s} 
        C_{x_b}^2 \sigma_{x_b \to f_i}^2  \Bigg).
		\label{BP_fv_var}	
        \end{align}
		\label{BP_fv_mean_var}%
		\end{subequations}   

After the factor node $f_i$ receives the messages from all of the neighbouring variable nodes from the set $\mathcal{X}_i\setminus x_s$, it evaluates the message $\mu_{f_i \to x_s}(x_s)$ according to \eqref{BP_fv_mean} and \eqref{BP_fv_var}, and sends it to the variable node $x_s$. 

\emph{Marginal inference:} It can be show that the marginal of the variable node $x_s$, illustrated in Fig. \ref{Fig_marginal}, is represented by the Gaussian function: 
\begin{equation}
        \begin{gathered}
        p(x_s) \propto 
        \mathcal{N}(\hat x_s|x_s,\sigma_{x_s}^2),
        \end{gathered}
		\label{BP_marginal_gauss}
		\end{equation} 
with the mean value $\hat x_s$ and variance $\sigma_{x_s}^2$:		
		\begin{subequations}
        \begin{align}
        \hat x_s &= 
        \Bigg( \sum_{f_c \in \mathcal{F}_s}
        \cfrac{z_{f_c \to x_s}}{\sigma_{f_c \to x_s}^2}\Bigg)
        \sigma_{x_s}^2
        \label{BP_marginal_mean} \\
		\cfrac{1}{\sigma_{x_s}^2} &= 
		\sum_{f_c \in \mathcal{F}_s}
		\cfrac{1}{\sigma_{f_c \to x_s}^2}.
		\label{BP_marginal_var}        
        \end{align}
        \label{BP_marginal_mean_var}		
		\end{subequations} 

Finally, the mean-value $\hat x_s$ is adopted as the estimated value of the state variable $x_s$.

\emph{Message scheduling:} The SE scenario is in general an instance of Loopy BP since the corresponding factor graph usually contains cycles. Loopy BP is an iterative algorithm and requires a message-passing schedule which, in this work, is selected as the usual synchronous schedule \cite{tsp_cosovic}.
 		
\section{Real-time State Estimation Using Belief Propagation}
In this section, we propose a fast and robust BP-based SE algorithm that can update the state estimate vector $\hat{\mathbf x}$ in a time-continuous process. Hence, the algorithm can handle each new measurement $M_i \in \mathcal{M}_{\mathrm{rt}}$ as soon as it is delivered from telemetry to the computational unit. Further, using the BP SE algorithm, it is possible to compute the state estimate vector $\hat{\mathbf x}$ without resorting to observability analysis.

The proposed SE solution is based on the fact that the BP-based algorithm is robust in terms of handling the ill-conditioned scenarios caused by significant differences between values of variances (e.g., PMU measurements and pseudo-measurements). Ideally, pseudo-measurements should not affect the solution within observable islands (i.e., determined with real-time measurements), therefore the variance of pseudo-measurements $M_i \in \mathcal{M}_{\mathrm{ps}}$ should be set to $\sigma_i^2 \to \infty$. In the conventional SE this concept is a source of ill-conditioned system. Hence, the values of pseudo-measurement variances should be defined to prevent ill-conditioned situations and ensure numerical stability of the SE algorithm (e.g., $10^{10}-10^{15}$). On the other hand, inability to define $\sigma_i^2 \to \infty$ causes that pseudo-measurements have more or less impact on the state estimate $\hat{\mathbf x}$, and thus the number of pseudo-measurements should be minimized to produce an observable system.

The BP SE algorithm allows the inclusion of an arbitrary number of pseudo-measurements with an extremely large values of variances (e.g., $10^{60}$), hence the impact on the observable island is negligible. Consequently, observable islands will have unique solution according to the real-time measurements, while unobservable islands will be determined according to both real-time and pseudo-measurements. Therefore, we propose a model where the network topology processor generates bus/branch model and assigns all possible measurements that exist in the power system, setting their variances to suitable values. 

\begin{figure}[ht]
	\centering
	\begin{tabular}{@{}c@{}}
	\subfloat[]{\label{Fig_bus_branch}
	\includegraphics[width=3.0cm]{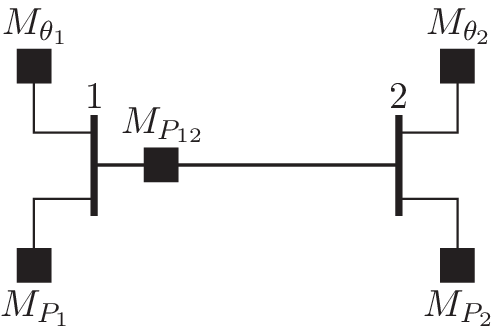}}
	\end{tabular}\quad
	\begin{tabular}{@{}c@{}}
	\subfloat[]{\label{Fig_DC_graph}
	\includegraphics[width=3.5cm]{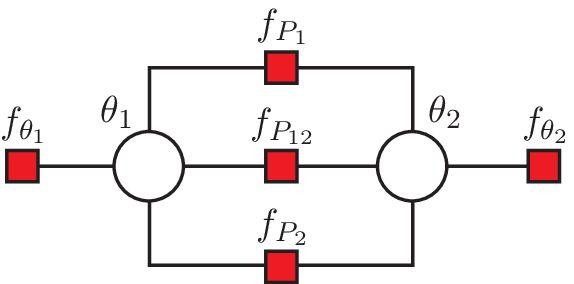}}
	\end{tabular}
	\caption{ Transformation of the bus/branch model and measurement configuration 
	(subfigure a) into the corresponding factor graph for the DC model 
	(subfigure b).}
	\label{Fig_DC}
	\end{figure}

Without loss of generality, we demonstrate this procedure by a toy-example, using a simple bus/branch model shown in Fig. \ref{Fig_bus_branch} where all the possible measurements are assigned. The first step is converting the bus/branch model and its measurements configuration into the corresponding factor graph illustrated in Fig. \ref{Fig_DC_graph}. 
We assume, for the time being, that all the measurements are pseudo-measurements 
$\mathcal{M} \equiv$  $\mathcal{M}_{\mathrm{ps}} =$ $\{M_{\theta_1},$ $M_{\theta_2},$ $M_{P_1},$ $M_{P_2},$ $M_{P_{12}}\}$ and $\mathcal{M}_{\mathrm{rt}} = \{\emptyset\}$, noting that the system is unobservable. Using equations \eqref{BP_vf_mean_var}, \eqref{BP_fv_mean_var} and \eqref{BP_marginal_mean_var} the BP algorithm will compute the state estimate vector $\hat{\mathbf x}$ according to the set of factor nodes $\mathcal{F}$ defined by the set of pseudo-measurements $\mathcal{M} \equiv$  $\mathcal{M}_{\mathrm{ps}}$. Hence, the system is defined according to the prior knowledge in lack of real-time measurements.

Subsequently, in an arbitrary moment, we assume that the computational unit received a real-time measurement $\mathcal{M}_{\mathrm{rt}} =$ $\{M_{\theta_1}\}$, which determines an observable island that contains bus $1$, while bus 2 remains within unobservable island. The BP algorithm in continuous process will compute the new value of state estimate $\hat{\theta}_1$ according to $M_{\theta_1}$, with insignificant impact of (high-variance) pseudo-measurements $\mathcal{M}_{\mathrm{ps}} \setminus \{M_{\theta_1}\}$, while the value of the state estimate $\hat{\theta}_2$ will be defined according to both $M_{\theta_1}$ and $\mathcal{M}_{\mathrm{ps}}\setminus \{M_{\theta_1}\}$.   

Assuming that subsequently, the computational unit receives an additional real-time measurement $M_{P_{12}}$, the system will be observable. The state estimate $\hat{\mathbf x}$ at that moment will be computed according to the real-time measurements $\mathcal{M}_{\mathrm{rt}} =$ $\{M_{\theta_1},$ $M_{P_{12}}\}$, with negligible influence of pseudo-measurements $\mathcal{M}_{\mathrm{ps}} \setminus \{M_{\theta_1}, M_{P_{12}}\}$.

Based on our extensive numerical analysis on large IEEE test cases, the proposed algorithm is able to track the state of the system in the continuous process without need for observability analysis. Note that, due the fact that the values of state variables usually fluctuate in narrow boundaries, in normal conditions, the continuous algorithm allows for fast response to new each measurement.     

\section{Numerical Results}      
\begin{figure}[ht]
	\centering
	\includegraphics[width=65mm]{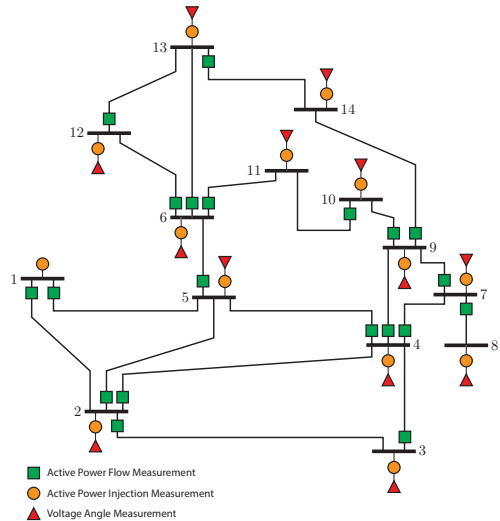}
	\caption{The IEEE 14 bus test case with measurement configuration.}
	\label{fig_IEEE14}
	\end{figure}

We evaluate the performance of the proposed algorithm using the IEEE 14 bus test case with the measurement configuration shown in Fig. \ref{fig_IEEE14}.  

\subsection{Simulation Setup}
We start by a given IEEE 14 bus test case and apply the DC power flow analysis to generate the exact solution for voltage angles and active powers across the network. Further, we corrupt the exact solution by
the additive white Gaussian noise of variance $\sigma^2$ to generate set of measurements $\mathcal{M}$. 

The slack bus is bus 1 where the voltage angle has a given value $\theta_1 = 0$, therefore, the variance is $\sigma_1^2 \to 0$ (e.g. we use $\sigma_1^2 = 10^{-60}\,\mathrm{deg}$). Throughout this section, the variance of active power flow and injection pseudo-measurements are $\sigma_{\mathrm{ps}}^2 = 10^{60}\,\mathrm{MW}$, while voltage angle pseudo-measurements have $\sigma_{\mathrm{ps}}^2 = 10^{60}\,\mathrm{deg}$. Note that the base power for the IEEE 14 bus test case is $100\,\mathrm{MVA}$. 

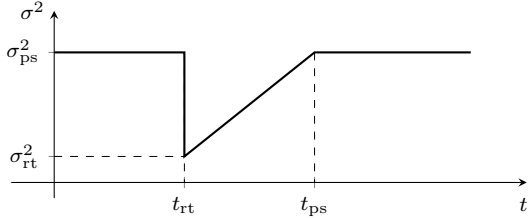
\begin{figure}[ht]
	\centering
	\begin{tikzpicture}
  	\begin{axis}[axis lines=center, axis equal image, enlargelimits=true,
  	y label style={at={(0.0,1.1)}}, x label style={at={(1.02,-0.12)}},
    xlabel={$t$},
	ylabel={$\sigma^2$},
    label style={font=\footnotesize},   	
    xtick={1, 2},
    xticklabels={$t_{\mathrm{rt}}$, $t_{\mathrm{ps}}$},
    ytick={0.2,1},
    yticklabels={$\sigma_{\mathrm{rt}}^2$, $\sigma_{\mathrm{ps}}^2$},
    tick label style={font=\footnotesize},
    ymin = 0, ymax = 1.2,   	
   	xmin = 0, xmax = 3.3]
   	\addplot [black, no markers, thick] coordinates {(0,1) (1,1) (1,0.2) (2,1) (3.2,1)};
   	\addplot [black, no markers, dashed, very thin] coordinates {(0,0.2) (1,0.2)}; 
   	\addplot [black, no markers, dashed, very thin] coordinates {(2,1) (2,0)};
   	\addplot [black, no markers, dashed, very thin] coordinates {(1,1) (1,0)};  
  	\end{axis}
	\end{tikzpicture}
	\caption{The time-dependent function of variances for real-time measurements.}
	\label{fig_var}
	\end{figure}

In each test case (described below), the algorithm starts at the time instant $t = 0$ initialized using the full set of pseudo-measurements $\mathcal{M} \equiv \mathcal{M}_{\mathrm{ps}}$ generated according to historical data.
Consider an arbitrary measurement $M_i \in \mathcal{M}$. This measurement is initialized as pseudo-measurement, i.e., at $t=0$, $M_i \in \mathcal{M}_{\mathrm{ps}}$. Let $t_{\mathrm{rt}}$ denotes the time instant when the computational unit has received the real-time measured value of $M_i$ with the predefined value of variance $\sigma_{\mathrm{rt}}^2$. We model the ``aging'' of the information provided by this measurement by the linear variance increase over time up to the time instant $t_{\mathrm{ps}}$ where it becomes equal to $\sigma_{\mathrm{ps}}^2$ (Fig \ref{fig_var}). In other words, we assume $M_i \in \mathcal{M}_{\mathrm{ps}}$ during $0 \leq t < t_{\mathrm{rt}}$ and $t \geq t_{\mathrm{ps}}$, while $M_i \in \mathcal{M}_{\mathrm{rt}}$ during $t_{\mathrm{rt}} \leq t < t_{\mathrm{ps}}$.
After the transition period $t \geq t_{\mathrm{ps}}$, $M_i$ is observed as pseudo-measurement until the next real-time measurement is received.
 
\subsection{Test Case 1}
In the following, we analyze performance of the proposed algorithm in the scenario characterized by significant differences between variances and observe influence of the pseudo-measurements on the state estimate $\hat{\mathbf x} \equiv \hat{\bm \uptheta}$.

In Table I, we define the (fixed) schedule and type of real-time measurements, where each real-time measurement is set to $\sigma_{\mathrm{rt}}^2 = 10^{-12}\,\mathrm{MW}$ at $t_{\mathrm{rt}}$ and we assume $t_{\mathrm{ps}} \to \infty$ (i.e., $\sigma_{\mathrm{rt}}^2$ remains at $10^{-12}\,\mathrm{MW}$ for $t > t_{\mathrm{rt}}$ ). 
The example is designed in such a way that, upon reception of each real-time measurement, due to its very low variance one of the states from the estimated state vector $\hat{\bm \uptheta}$ becomes approximately equal to the power flow solution.
	\begin{table}[ht]
	\centering
	\caption{Schedule and type of real-time measurements}
	\label{Tab1}
	\resizebox{\columnwidth}{!}{%
	\begin{tabular}{c|cc||c|cc}
	\hline
	\multicolumn{1}{c|}{Time} & \multicolumn{2}{c||}{Active power flow $M_{P_{ij}}$} & 
	\multicolumn{1}{c|}{Time} & 
	\multicolumn{2}{c}{Active power flow $M_{P_{ij}}$} \rule{0pt}{1ex}\rule{0pt}{3ex}\\
	$t_{\mathrm{rt}} (\mathrm{s})$ & \multicolumn{1}{c}{from bus $i$}  & \multicolumn{1}{c||}{to bus $j$} & 
	$t_{\mathrm{rt}} (\mathrm{s})$ & \multicolumn{1}{c}{from bus $i$}  & \multicolumn{1}{c}{to bus $j$}
	\rule{0pt}{3ex}\\
	\hline
	1     & 1 & 2  & 8 & 7 & 9   
	\rule{0pt}{3ex}\\
	2     & 2 & 3  & 9 & 9 & 10    
	\rule{0pt}{2ex}\\
	3     & 3 & 4  & 10 & 10 & 11   
	\rule{0pt}{2ex}\\
	4     & 4 & 5  & 11 & 6 & 12  
	\rule{0pt}{2ex}\\
	5     & 5 & 6  & 12 & 12 & 13  
	\rule{0pt}{2ex}\\
	6     & 4 & 7  & 13 & 13 & 14  
	\rule{0pt}{2ex}\\
	7     & 7 & 8  &  &  &  
	\rule{0pt}{2ex}\\							
	\hline
	\end{tabular}}
	\end{table}

	\begin{figure}[ht]
	\centering
	\begin{tikzpicture}
  	\begin{axis}[xmajorticks=false,width=8cm,height=3.0cm,at={(0cm,0cm)},
   	y tick label style={/pgf/number format/.cd,fixed,
   	fixed zerofill, precision=1, /tikz/.cd},
   	x tick label style={/pgf/number format/.cd,
   	set thousands separator={},fixed},
   	legend cell align=left,
   	legend style={at={(0.6,0.69)},anchor=west,font=\tiny},
   	legend entries={Power Flow Solution, BP SE Solution},
   	xlabel={},
   	ylabel={$\theta_3$ (deg)},
   	label style={font=\scriptsize},   	
   	grid=major,
    xtick={1,2,3,4,5,6,7,8,9,10,11,12,13,14},
   	ytick={-6, -9.5, -13},
   	tick label style={font=\scriptsize},
   	ymin = -14.0, ymax = -5,   	
   	xmin = 0, xmax = 14]
   	\addplot [red, no markers] coordinates {(0,-12.9536631292105) (14,-12.9536631292105)}; 
   	\addplot[blue] 
   	table [x={time}, y={T3}] {case_1.txt};
  	\end{axis}
  
  	
  	\begin{axis}[xmajorticks=false,width=8cm,height=3.0cm,at={(0cm,-1.6cm)},
   	y tick label style={/pgf/number format/.cd,fixed,
   	fixed zerofill, precision=1, /tikz/.cd},
   	x tick label style={/pgf/number format/.cd,
   	set thousands separator={},fixed},
   	ylabel={$\theta_8$ (deg)},
   	label style={font=\scriptsize},   	
   	grid=major,
    xtick={1,2,3,4,5,6,7,8,9,10,11,12,13,14},
    ytick={-14, -18, -22},
   	tick label style={font=\scriptsize},
   	ymin = -23.0, ymax = -13,   	
   	xmin = 0, xmax = 14]
   	\addplot [red, no markers] coordinates {(0,-13.9070545899204) (14,-13.9070545899204)}; 
   	\addplot[blue]
   	table [x={time}, y={T8}] {case_1.txt};
  	\end{axis} 
%
  	
	
    \begin{axis}[width=8cm,height=3.0cm,at={(0cm,-3.2cm)},
   	y tick label style={/pgf/number format/.cd,fixed,
   	fixed zerofill, precision=1, /tikz/.cd},
   	x tick label style={/pgf/number format/.cd,
   	set thousands separator={},fixed},
   	xlabel={Time (s)},
   	ylabel={$\theta_{14}$ (deg)},
   	label style={font=\scriptsize},   	
   	grid=major,
    xtick={1,2,3,4,5,6,7,8,9,10,11,12,13,14},
  	ytick={-16, -19.5, -23},
   	tick label style={font=\scriptsize},
    ymin = -24, ymax = -15,   	
   	xmin = 0, xmax = 14]
   	\addplot [red, no markers] coordinates {(0,-17.1882875702935) (14,-17.1882875702935)}; 
   	\addplot[blue]
   	table [x={time}, y={T14}] {case_1.txt};
  	\end{axis}
	\end{tikzpicture}
	\caption{Real-Time estimates of voltage angles $\theta_3$, 
	$\theta_8$ and $\theta_{14}$ where the computational unit received 
	active power flow real-time measurements every $t = 1\,\mathrm{s}$ with variance
	$\sigma_{\mathrm{rt}}^2 = 10^{-12}\,\mathrm{MW}$.}
	\label{Fig_case1}
	\end{figure}
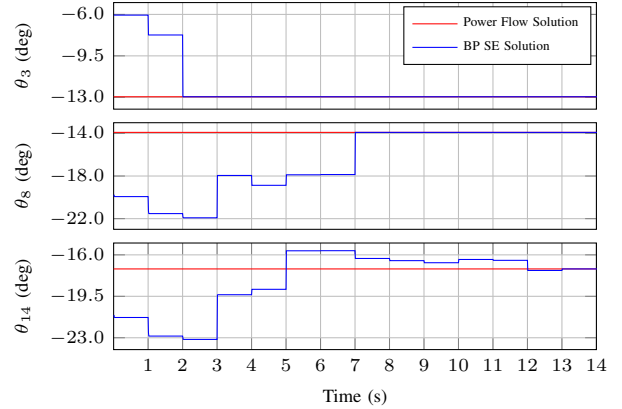

Fig. \ref{Fig_case1} shows estimated values of voltage angles $\theta_3$, $\theta_8$ and $\theta_{14}$ for the scenario defined in Table \ref{Tab1}. One can note the robustness of the proposed BP SE solution in a sense that, at any time instant, the extreme difference in variances between already received real-time measurements and remaining set of pseudo-measurements (that typically lead to ill-conditioned scenarios), are accurately solved by the BP estimator. As expected, in our pre-designed example, we clearly note a sequential refinement of the state estimate, where each new received real-time measurement $M_{P_{ij}}$ accurately defines the corresponding state variable $\theta_j$. More precisely, starting from the slack bus that has a known state value,  the real-time measurement $M_{P_{12}}$ specifies the state value of $\theta_2$ at time $t=1\,\mathrm{s}$. The chain of refinements repeats successively until $t=13\,\mathrm{s}$ when the final state variable $\theta_{14}$ is accurately estimated.   

Although somewhat trivial, the above example demonstrates that the BP-based SE algorithm provides a solution according to the real-time measurements, irrespective of the presence of (all) pseudo-measurements. In addition, Fig. \ref{Fig_case1} shows how BP influence propagates through the network (e.g., upon reception, measurement $M_{P_{12}}$ affects the distant state variable $\theta_{14}$).
 
\subsection{Test Case 2}

In order to investigate how fast BP influence propagates through the network, we use the same setup given in Section IV-B, and analyse the response of the system to the received real-time measurement of different variance $\sigma_{\mathrm{rt}}^2 =$ $\{20^2,$ $10^2,$ $10^{-2}\}\,\mathrm{MW}$. In particular, we track the convergence of the (iterative message passing) BP algorithm over time, from the moment the real-time measurement is received, to the moment when the state estimate reaches a steady state.

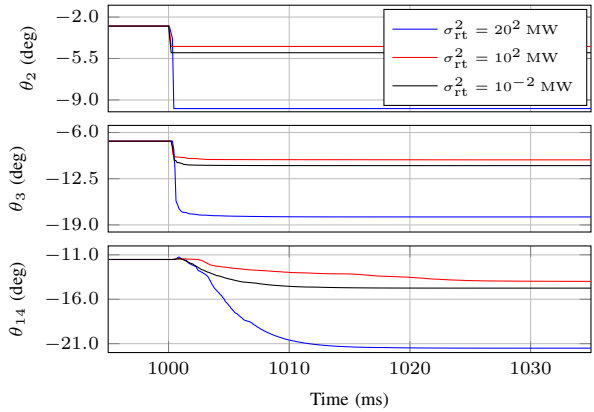
\begin{figure}[ht]
	\centering
	\begin{tikzpicture}
    \begin{axis}[xmajorticks=false,width=8cm,height=3cm,at={(0cm,0cm)},
   	y tick label style={/pgf/number format/.cd,fixed,
   	fixed zerofill, precision=1, /tikz/.cd},
   	x tick label style={/pgf/number format/.cd,
   	set thousands separator={},fixed, precision=1},
   	legend cell align=left,
    legend style={at={(0.57,0.51)},anchor=west,font=\tiny},
   	legend entries={$\sigma_{\mathrm{rt}}^2=20^2\,\mathrm{MW}$, 
   	$\sigma_{\mathrm{rt}}^2=10^2\,\mathrm{MW}$,
   	$\sigma_{\mathrm{rt}}^2=10^{-2}\,\mathrm{MW}$},
   	ylabel={$\theta_{2}$ (deg)},
   	label style={font=\scriptsize},   	
   	grid=major,
  	ytick={-9, -5.5, -2.0},
   	tick label style={font=\scriptsize},
 	ymin = -10, ymax = -1,   	
 	xmin = 995, xmax = 1035]
   	\addplot[blue]
   	table [x={time}, y={T2}] {case_n20.txt};
   	\addplot[red]
   	table [x={time}, y={T2}] {case_n10.txt};
   	\addplot[black]
   	table [x={time}, y={T2}] {case_n10-1.txt};
  	\end{axis}
  	
  	\begin{axis}[xmajorticks=false,width=8cm,height=3.0cm,at={(0cm,-1.6cm)},
   	y tick label style={/pgf/number format/.cd,fixed,
   	fixed zerofill, precision=1, /tikz/.cd},
   	x tick label style={/pgf/number format/.cd,
   	set thousands separator={},fixed},
   	ylabel={$\theta_{3}$ (deg)},
   	label style={font=\scriptsize},   	
   	grid=major,
  	ytick={-19, -12.5, -6.0},
   	tick label style={font=\scriptsize},
 	ymin = -20, ymax = -5,   	
 	xmin = 995, xmax = 1035]
   	\addplot[blue]
   	table [x={time}, y={T3}] {case_n20.txt};
   	\addplot[red]
   	table [x={time}, y={T3}] {case_n10.txt};
   	\addplot[black]
   	table [x={time}, y={T3}] {case_n10-1.txt};
  	\end{axis}
  	
 
    \begin{axis}[width=8cm,height=3cm,at={(0cm,-3.2cm)},
   	y tick label style={/pgf/number format/.cd,fixed,
   	fixed zerofill, precision=1, /tikz/.cd},
   	x tick label style={/pgf/number format/.cd,
   	set thousands separator={},fixed, precision=1},
   	ylabel={$\theta_{14}$ (deg)},
   	xlabel={Time (ms)},   	
   	label style={font=\scriptsize},   	
   	grid=major,
  	ytick={-21, -16, -11},
   	tick label style={font=\scriptsize},
 	ymin = -22, ymax = -10,   	
 	xmin = 995, xmax = 1035]
   	\addplot[blue]
   	table [x={time}, y={T14}] {case_n20.txt};
   	\addplot[red]
   	table [x={time}, y={T14}] {case_n10.txt};
   	\addplot[black]
   	table [x={time}, y={T14}] {case_n10-1.txt};
  	\end{axis} 
	\end{tikzpicture}
	\caption{Real-Time estimates of voltage angles $\theta_2$, $\theta_3$  
	and $\theta_{14}$ where the computational unit received 
	active power flow real-time measurement $M_{P_{12}}$ at the time 
	$t = 1\,\mathrm{s}$ with variances
	$\sigma_{\mathrm{rt}}^2 = \{20^2, 10^2, 10^{-2}\}\,\mathrm{MW}$.}
	\label{Fig_case2}
	\end{figure}

Fig. \ref{Fig_case2} illustrates the influence of the real-time measurement $M_{P_{12}}$ received at $t_{\mathrm{rs}} = 1\,\mathrm{s}$, on the state variables  $\theta_2$, $\theta_3$ and $\theta_{14}$. As expected, the received real-time measurement has almost immediate impact on the state variable $\theta_2$, where steady state occurs within $t < 1\,\mathrm{ms}$, even for the high value of measurement variance $\sigma_{\mathrm{rt}}^2=20^2\,\mathrm{MW}$. Further, this real-time measurement will influence the entire system through iterative BP message exchanges. As expected, increasing the distance between the measurement location and the bus location, more time is need for the corresponding state variable to reach the steady state. For example, steady state of the state variable $\theta_{14}$ occurs within $t < 25\,\mathrm{ms}$. 

To summarize, the algorithm is able to provide fast response on the received real-time measurements and, for the DC SE framework, it is able to support both WAMS and SCADA technology in terms of the required computational delays\footnote{Our ongoing work aims to extend these results to the more complex BP-based AC SE model \cite{tsp_cosovic}.}. 
	
\vspace{-0.35cm}
\subsection{Test Case 3}
	\begin{figure}[ht]
	\centering
	\begin{tikzpicture}
  	\begin{axis}[xmajorticks=false,width=8cm,height=3.1cm,at={(0cm,0cm)},
   	y tick label style={/pgf/number format/.cd,fixed,
   	fixed zerofill, precision=1, /tikz/.cd},
   	x tick label style={/pgf/number format/.cd,
   	set thousands separator={},fixed},
   	legend cell align=left,
   	legend style={at={(0.704,0.709)},anchor=west,font=\tiny},
   	legend entries={Power Flow, BP SE},
   	xlabel={},
   	ylabel={$\theta_3$ (deg)},
   	label style={font=\scriptsize},   	
   	grid=major,
    xtick={50, 100, 150, 200, 250, 300},
   	ytick={-7, -12, -17},
   	tick label style={font=\scriptsize},
   	ymin = -19.0, ymax = -5,   	
   	xmin = 0, xmax = 300]
   	\addplot [red, no markers] coordinates {(0,-12.9536631292105) (100,-12.9536631292105) 		
   	(100,-7.21648794584402) (200,-7.21648794584402) (200,-16.4004583212321) (300,-16.4004583212321)};
   	\addplot[blue] 
   	table [x={time}, y={T3}] {case_2.txt};
  	\end{axis}
  
  	
  	\begin{axis}[xmajorticks=false,width=8cm,height=3.0cm,at={(0cm,-1.6cm)},
   	y tick label style={/pgf/number format/.cd,fixed,
   	fixed zerofill, precision=1, /tikz/.cd},
   	x tick label style={/pgf/number format/.cd,
   	set thousands separator={},fixed},
   	ylabel={$\theta_8$ (deg)},
   	label style={font=\scriptsize},   	
   	grid=major,
    xtick={50, 100, 150, 200, 250, 300},
   	ytick={-2, -12, -22},
   	tick label style={font=\scriptsize},
   	ymin = -24, ymax = 0,   	
   	xmin = 0, xmax = 300]
   	\addplot [red, no markers] coordinates {(0,-13.9070545899204) (100,-13.9070545899204) 		
   	(100,-9.57566308601556) (200,-9.57566308601556) (200,-18.8061641931343) (300,-18.8061641931343)};
  	\addplot[blue]
   	table [x={time}, y={T8}] {case_2.txt};
  	\end{axis} 
%
  	
	
    \begin{axis}[width=8cm,height=3.0cm,at={(0cm,-3.2cm)},
   	y tick label style={/pgf/number format/.cd,fixed,
   	fixed zerofill, precision=1, /tikz/.cd},
   	x tick label style={/pgf/number format/.cd,
   	set thousands separator={},fixed},
   	xlabel={Time (s)},
   	ylabel={$\theta_{14}$ (deg)},
   	label style={font=\scriptsize},   	
   	grid=major,
    xtick={50, 100, 150, 200, 250, 300},
  	ytick={-2, -15, -28},
   	tick label style={font=\scriptsize},
  	ymin = -30, ymax = 0,   	
   	xmin = 0, xmax = 300]
   	\addplot [red, no markers] coordinates {(0,-17.1882875702935) (100,-17.1882875702935) 		
   	(100,-11.5246130227784) (200,-11.5246130227784) (200,-21.1725201082984) (300,-21.1725201082984)}; 
   	\addplot[blue]
   	table [x={time}, y={T14}] {case_2.txt};
  	\end{axis}
	\end{tikzpicture}
	\caption{Real-time estimates of voltage angles $\theta_3$, $\theta_8$ 
	and $\theta_{14}$ where real-time measurements arrived at the 
	computational unit according to Poisson process.}
	\label{Fig_case3}
	\end{figure}
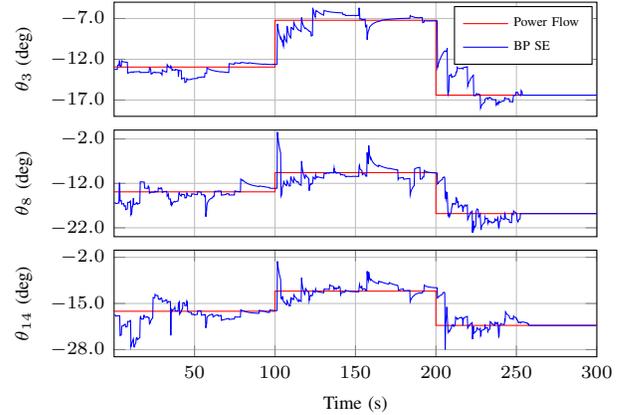
 
In the final scenario, we consider the dynamic scenario in which the power system changes values of both generations and loads every $100\,\mathrm{s}$. In the interval between $t=0$ and $t=250\,\mathrm{s}$, only active power flow and injection real-time measurements are available with variances $\sigma_{\mathrm{rt}}^2 = 10^{2}\,\mathrm{MW}$ and  $t_{\mathrm{ps}}-t_{\mathrm{rt}} = 10^{3}\,\mathrm{s}$.\footnote{Although the period of $10^{3}\,\mathrm{s}$ may appear large, note that this is compensated by very high variance $\sigma_{\mathrm{ps}}^2 = 10^{60}\,\mathrm{MW}$ at $t_{\mathrm{ps}}$.} After $250\,\mathrm{s}$, the voltage angle real-time measurements become available with parameters $\sigma_{\mathrm{rt}}^2 = 10^{-6}\,\mathrm{deg}$ and $t_{\mathrm{ps}} \to \infty$. For every measurement, arrival process in each interval is modeled using Poisson process with average inter-arrival time $1/\lambda$, where for active power flow and injection real-time measurements we set $\lambda = 0.05$ and for angle real-time measurements $\lambda = 0.5$.

Fig. \ref{Fig_case3} shows state estimates of state variables $\theta_3$, $\theta_8$ and $\theta_{14}$ over the time interval of $300\,\mathrm{s}$ for the described scenario. During the first $250\,\mathrm{s}$, the BP SE provides state estimates according to incoming noisy real-time measurements and, as apparent from the figure, each new real-time measurement will affect the current state of the system. After $t=250\,\mathrm{s}$, the voltage angle real-time measurements arrived with constant and very low variance, thus providing state estimates which are considerably more accurate.

\section{Conclusions}
We presented the fast real-time DC SE model based on the powerful BP algorithm, which is able to provide state estimates without resorting to observability analysis. The proposed BP estimator can be distributed and parallelized which allows for flexible and low-delay centralized or distributed implementation suitable for integration in emerging WAMS. For the future work, we plan to provide extensive numerical analysis of the proposed algorithm, including the AC SE model implemented within the same framework, and extended to the generalized SE model.

\bibliographystyle{IEEEtran}
\bibliography{smartgrid}
\end{document}